\documentclass[journal,twoside]{IEEEtran}
\usepackage[margin=0.7in]{geometry}

\usepackage[T1]{fontenc}

\usepackage{cite}
\usepackage[pdftex]{graphicx}
\graphicspath{{Figure/}}
\usepackage[cmex10]{amsmath}
\usepackage{amsthm}
\usepackage{amssymb}
\usepackage{algorithmic}
\usepackage{array}
\usepackage{color}
\usepackage{epstopdf}
\usepackage{amsfonts}

\usepackage{pgfplots}
\pgfplotsset{compat=1.3}
\usepackage{tikz}							
\usetikzlibrary{shapes}
\usetikzlibrary{spy}
\usetikzlibrary{circuits}
\usetikzlibrary{arrows}
\usetikzlibrary{spy}
\usepackage{subfig}

\usepackage[ruled,vlined]{algorithm2e}

\newcommand{\vect}[1]{\boldsymbol{\mathrm{#1}}}
\newcommand{\mat}[1]{\boldsymbol{\mathrm{#1}}}

\newcommand{\tr}{\mathrm{tr}}

\newcommand{\diag}{\text{diag}}

\usepackage{mathtools}

\theoremstyle{remark}
\newtheorem{theorem}{Theorem}

\begin{document}
\linespread{1}
\title{Secrecy Capacity of FBMC-OQAM Modulation\\
over Frequency Selective Channel}

%
%

\author{Fran\c{c}ois Rottenberg,~Philippe De Doncker,~François Horlin and Jérôme Louveaux
	\thanks{The work of F. Rottenberg was supported by the Belgian National Science Foundation (FRS-FNRS). F. Rottenberg and J. Louveaux are with the Université catholique de Louvain, 1348 Louvain-la-Neuve, Belgium. F. Rottenberg, P. D. Doncker and F. Horlin are with the Université libre de Bruxelles, 1050 Brussels, Belgium.}
\vspace{-1em}
}

\maketitle

\begin{abstract}
	This paper studies the information-theoretic secrecy capacity of an Offset-QAM-based filterbank multicarrier (FBMC-OQAM) communication over a wiretap frequency selective channel. The secrecy capacity is formulated as an optimization problem which has a closed-form solution in the high signal-to-noise ratio (SNR) regime. Two of the most common equalization strategies in FBMC-OQAM are considered, namely, single-tap and multi-tap equalization. For the sake of comparison, we also consider the secrecy capacity of a generic modulation and a cyclic prefix-orthogonal frequency division multiplexing (CP-OFDM) modulation. As a result, we find that FBMC-OQAM is particularly competitive for medium-to-long burst transmissions.
\end{abstract}


\section{Introduction}\label{section:Introduction}
\linespread{1}

The information-theoretic secrecy-capacity is defined as the number of bits per channel use that can be reliably transmitted from a legitimate transmitter (Alice) to a legitimate receiver (Bob) while guaranteeing a negligible information leakage to the eavesdropper (Eve). The seminal work of Wyner \cite{Wyner1975} and its extension to more general channels \cite{Csiszar1978}, have shown that a "physical advantage" at Bob with respect to Eve is required to guarantee a larger-than-zero secrecy capacity. Multipath channels lead to channel frequency selectivity. If Alice knows the channel of Bob and Eve, she can modulate her signal to take benefit of frequency bins where Bob's channel has an advantage over Eve's channel. This scenario has been studied in details in the case of multicarrier modulations, including CP-OFDM \cite{Jorswieck2008,li2009secrecy,Renna2012}.

However, to the best of the authors knowledge, we are the first to analyze the secrecy capacity of the FBMC-OQAM modulation, which has received increasing attention in the last decades as an attractive alternative to CP-OFDM modulation \cite{Boroujeny2011}. In this paper, we consider that Alice and Bob communicate over a frequency selective channel using FBMC-OQAM modulation/demodulation while Eve tries to eavesdrop on the conversation. Based on this model, we formulate the secrecy capacity as an optimization problem that can be solved in closed-form at high SNR. At Bob side, two of the most common equalization strategies in FBMC-OQAM are considered, namely, multi-tap and single-tap equalization \cite{rottenberg2018fbmc}. We demonstrate that both equalization schemes lead to equivalent performance for mildly frequency selective channels. At Eve side, we additionally consider the loss in secrecy occuring if she is not constrained to apply conventional FBMC-OQAM demodulation. For the sake of comparison, we also consider the secrecy capacity of a generic modulation and a CP-OFDM modulation. In the end, we show that FBMC-OQAM is particularly competitive for medium-to-long burst transmissions.

\textbf{Notations}: 
Vectors and matrices are denoted by bold lowercase and uppercase letters $\vect{a}$ and $\mat{A}$, respectively (resp.). Superscripts $^*$, $^T$, $^H$ and $^{\dagger}$ stand for conjugate, transpose, Hermitian transpose and Moore-Penrose pseudo-inverse. The symbols $\tr[.]$, $\mathbb{E}(.)$, $\Im(.)$ and $\Re(.)$ denote the trace, expectation, imaginary and real parts, respectively. $\jmath$ is the imaginary unit. $\|\mat{A}\|$ and $|\mat{A}|$ are the Frobenius norm and determinant respectively. $\mat{I}_N$ denotes the identity matrix of order $N$. $\mat{0}_{N\times M}$ is a zero matrix of size $N\times M$. Subscripts of matrices are dropped whenever matrix dimensions are clear from the context. 
$\diag(\vect{a})$ returns a diagonal matrix with $\vect{a}$ on its diagonal. The positive part of a real quantity is denoted by $[a]^+=\max(a,0)$. $\sigma_n(\mat{A})$ (or $\lambda_n(\mat{A})$) is the $n$-th largest singular (or eigenvalue) of $\mat{A}$. $\otimes$ stands for the Kronecker product.

\section{System Model}
\label{section:system_model}

\begin{figure}[!t]  
	\centering
	
	\resizebox{0.45\textwidth}{!}{%
		{\includegraphics[clip, trim=0cm 16cm 23cm 0cm, scale=1]{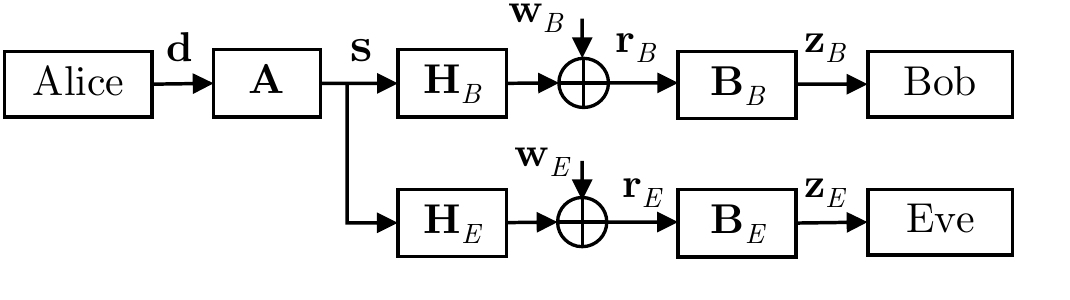}} 
	}
	\vspace{-1em}
	\caption{Transmission model over multipath channel equivalent to a special case of MIMO wiretap channel.}
	\label{fig:transmission_model}
	\vspace{-2em}
\end{figure}

As depicted in Fig.~\ref{fig:transmission_model}, we consider a conventional wiretap channel where Alice wants to communicate secretly with Bob while Eve tries to eavesdrop on the communication. Alice transmits a group of symbols $\vect{d}$, modulated with matrix $\mat{A}$ as $\vect{s}=\mat{A}\vect{d}$ where $\vect{s}\in \mathbb{C}^{N\times 1}$. We consider a constraint on the power of the transmitted signal so that $\tr\left[\mat{R}_s\right]\leq P$ with $\mat{R}_s=\mathbb{E}(\vect{s}\vect{s}^H) \in \mathbb{C}^{N\times N}$. The channel is modeled as a multipath channel that is considered quasi-static, \textit{i.e.}, it remains constant over the duration of the $N$ symbols. {\color{black}The channel impulse responses from Alice to Bob and Eve are denoted by $h_B[n]$ and $h_E[n]$ respectively, of length $L_B$ and $L_E$, and are assumed to be known by Alice, which is a common assumption \cite{bloch2011physical}. The knowledge of Bob's channel can be acquired in practice through pilots and feedback, or based on channel reciprocity in time division duplexing mode. On the other hand, the knowledge of Eve's channel is a stronger assumption as she might remain passive. However, if she is an active network node, Alice may also have knowledge of her channel in the same way as Bob.} The vector of received samples at Bob and Eve are denoted by $\vect{r}_B\in \mathbb{C}^{N+L_B-1\times 1}$ and $\vect{r}_E\in \mathbb{C}^{N+L_E-1\times 1}$. The multipath channel can be modeled as a special case of the multiple-input-multiple-output (MIMO) wiretap channel \cite{kobayashi2009secured}
\begin{align}
	\vect{r}_B&=\mat{H}_B \vect{s} +\vect{w}_B,\ \vect{r}_E=\mat{H}_E \vect{s} +\vect{w}_E, \label{eq:basic_transmission_model}
\end{align}
where the "MIMO" channel matrix $\mat{H}_B\in \mathbb{C}^{N+L_B-1\times N}$ has a Toeplitz structure with vector $(h_B[0],\hdots, h_B[L_B-1],\vect{0}_{1\times N-1})^T$ as its first column and analogously for $\mat{H}_E\in \mathbb{C}^{N+L_E-1\times N}$. Note that $\mat{H}_B$ and $\mat{H}_E$ are both full rank by construction. The noise vectors $\vect{w}_B \in \mathbb{C}^{N+L_B-1\times 1}$ and $\vect{w}_E \in \mathbb{C}^{N+L_E-1\times 1}$ are modeled as zero mean circularly-symmetric complex Gaussian (ZMCSCG) vector with covariance $\mat{I}_{N+L_B-1}$ and $\mat{I}_{N+L_E-1}$ respectively. The normalization choice of a unit noise variance simplifies exposition and is not a loss of generality since a difference in SNR can be captured through a scaled channel gain. Bob (resp. Eve) applies matrix $\mat{B}_B$ (resp. $\mat{B}_E$) to demodulate and equalize the received samples, obtaining $\vect{z}_B$ (resp. $\vect{z}_E$).

\section{Generic Secrecy Capacity}

In this section, we are interested in the secrecy capacity without imposing any constraint on the transmitter and the receiver implying that $\mat{A}$, $\mat{B}_B$ and $\mat{B}_E$ are identity matrices. Using the general result of \cite{Csiszar1978}, the secrecy capacity of the MIMO wiretap model of (\ref{eq:basic_transmission_model}), under the power constraint $\tr\left[\mat{R}_s\right]\leq P$, can be written as
\begin{align} \label{eq:general_expression_capacity}
C_s=\max_{p(\vect{s},v),\tr\left[\mat{R}_s\right]\leq P} I(v;\vect{r}_B)-I(v;\vect{r}_E),
\end{align}
where $I(.;.)$ is the mutual information, $v$ is an auxiliary random variable defined such that $p(\vect{r}_B,\vect{r}_E|\vect{s},v)=p(\vect{r}_B,\vect{r}_E|\vect{s})$ with $p(.)$ being the probability density function. The result of \cite{Oggier2011} states that the secrecy capacity of the general Gaussian MIMO wiretap channel (\ref{eq:basic_transmission_model}) is attained without channel prefixing ($v=\vect{s}$) and $\vect{s}$ has to follow a ZMCSCG with covariance matrix $\mat{R}_s$. Using this result, (\ref{eq:general_expression_capacity}) can be further detailed so that the secrecy capacity and the covariance matrix $\mat{R}_s$ can be obtained by maximizing
\begin{align}
	C_s
	&=\max_{\mat{R}_s,\tr\left[\mat{R}_s\right]\leq P} \log \frac{\left| \mat{I} + \mat{H}_B \mat{R}_s \mat{H}_B^H \right|}{\left| \mat{I} + \mat{H}_E \mat{R}_s \mat{H}_E^H \right|}.\label{eq:secrecy_capacity}
\end{align}
This maximization problem is well known to be non convex and no general closed-form solution exists \cite{Khisti2007}. Still, in the high SNR regime, a closed-form solution of $C_s$ exists. The generalized singular value decomposition of the matrix pair $(\mat{H}_B,\mat{H}_E)$ allows us to decompose the MIMO wiretap channel into a set of parallel independent Gaussian wiretap channels. This is analogous to the case with no eavesdropper, where classical singular value decomposition allows to convert the MIMO channel in a set of independent parallel channels. Using the result of \cite[Th.~2]{Khisti2007} in the case of $\text{Ker}(\mat{H}_E)\cap \text{Ker}(\mat{H}_B)^\perp=\varnothing$ (since $\mat{H}_B$ and $\mat{H}_E$ are of full rank), we find
\begin{align*}
	\lim_{P\rightarrow +\infty} C_s &= \sum_{n=1}^N \left[\log \sigma_n^2\left(  \mat{H}_B \mat{H}_E^{\dagger} \right) \right]^+ 
	= \sum_{n=1}^N \left[\log \lambda_n\left(  \mat{C}_N \right) \right],
\end{align*}
where $\mat{C}_N=\mat{H}_B^H\mat{H}_B (\mat{H}_E^H\mat{H}_E)^{-1}$. {\color{black}The fact that $\text{Ker}(\mat{H}_E)\cap \text{Ker}(\mat{H}_B)^\perp=\varnothing$ directly implies that Alice cannot communicate secretly with Bob by modulating her signal such that it lies in the null space of Eve. Still, Alice can transmit in channel modes where the gain at Bob is higher than Eve but the secrecy capacity remains bounded as the SNR grows large. This in contrast with the multi-antenna case where Alice can use spatial beamforming to transmit in the null space of Eve so that the capacity can be unbounded as the SNR grows large \cite{Khisti2007}.}

Using the asymptotic properties of Toeplitz matrices, the authors in \cite{Renna2012} have further characterized the limiting behavior of the latest expression as the number of symbols $N$ in $\vect{s}$ grows large
\begin{align*}
	\lim_{N \rightarrow +\infty} \frac{1}{N}\sum_{n=1}^N \left[\log \lambda_n\left( \mat{C}_N \right) \right]^+=\int_{0}^{1} \left[\log \frac{|H_B(f)|^2}{|H_E(f)|^2} \right]^+ df,
\end{align*}
with $H_{B}(f)=\sum_{n=0}^{L_B-1} h_B[n]e^{-\jmath 2\pi f n}$ and $H_{E}(f)=\sum_{n=0}^{L_E-1} h_E[n]e^{-\jmath 2\pi f n}$.

\section{Secrecy Capacity of CP-OFDM}

This section details the CP-OFDM secrecy capacity, as was done in \cite{Renna2012}. We assume that the cyclic prefix (CP) length $L_{CP}$ is larger than $L_B-1$ and $L_E-1$. Under that condition, successive OFDM blocks do not interfere and it is sufficient to consider a single OFDM symbol that we denote by $\vect{d}\in \mathbb{C}^{M\times 1}$ where $M=N-L_{CP}$ is the number of subcarriers. The CP-OFDM modulation imposes the following structure on the transmitted symbols
\begin{align*}
	\vect{s}&=\mat{A}\vect{d},\
	\mat{A}=\begin{pmatrix}
	\mat{0}\ \mat{I}_{L_{CP}}\\
	\mat{I}_M \
	\end{pmatrix}\mat{F}^H,
\end{align*}
where $\mat{F}\in \mathbb{C}^{M \times M}$ is the unitary FFT matrix. Note that the total power is not equal before/after CP insertion so that the transmit power constraint becomes $\tr\left[\mat{R}_s\right]=\tr[\mat{A}\mat{R}_d\mat{A}^H]\leq P$. At the receiver, Bob uses conventional OFDM demodulation, \textit{i.e.}, CP removal and FFT
\begin{align*}
	\vect{z}_B&= \mat{B}_B\vect{r}_B= \mat{D}_B \vect{d}+ \tilde{\vect{w}}_B,
\end{align*}
with
\begin{align*}
	\mat{B}_B&= \mat{F}\begin{pmatrix}
	\mat{0}_{M \times L_{CP}}\ \mat{I}_M \ \mat{0}_{M \times L_B-1}
	\end{pmatrix}\\
	\mat{D}_B&= \mat{B}_B \mat{H}_B \mat{A} =\diag\left(H_B(f_0),...,H_B(f_{M-1})\right)\\
	\tilde{\vect{w}}_B&= \mat{B}_B\vect{w}_B,
\end{align*}
and $f_m=m/M$. At Eve side, two cases can be distinguished for its receiver structure $\mat{B}_E$: 1) Eve uses a generic (ideal) receiver in which case $\mat{B}_E=\mat{I}$ or 2) Eve uses conventional (sub-optimal) OFDM demodulation as Bob implying that $\mat{B}_E=\mat{B}_B$.
It was shown in \cite{Renna2012} that the secrecy capacity is attained if symbols $\vect{d}$ are ZMCSCG. Then, the secrecy capacity can be written as
\begin{align*}
	C_s^{\mathrm{OFDM}}&=\max_{\mat{R},\tr\left[\mat{R}\right]\leq P} \log \frac{\left| \mat{I} + \mat{T}_B \mat{R} \mat{T}_B^H \right|}{\left| \mat{I} + \mat{T}_E \mat{R} \mat{T}_E^H \right|},
\end{align*}
where $\mat{T}_B=\mat{D}_B\mat{C}^{-1/2}$ and $\mat{T}_E=\mat{B}_E \mat{H}_E \mat{A}\mat{C}^{-1/2}$ with $\mat{C}=\mat{A}^H\mat{A}$. The optimal input covariance matrix $\mat{R}_d=\mathbb{E}(\vect{d}\vect{d}^H)$ is related to matrix $\mat{R}$ as $\mat{R}_d=\mat{C}^{-1/2}\mat{R}\mat{C}^{-1/2}$. Using again the result of \cite[Th.~2]{Khisti2007}, the secrecy capacity at high SNR becomes
\begin{align*}
&\lim_{P\rightarrow +\infty} C_s^{\mathrm{OFDM}} = \sum_{m=1}^M \left[\log \sigma_m^2\left( \mat{T}_B \mat{T}_E^{\dagger} \right) \right]^+. 
\end{align*}
If Eve uses a conventional CP-OFDM receiver ($\mat{B}_E=\mat{B}_B$), we have $\mat{T}_E=\mat{D}_E\mat{C}^{-1/2}$ with $\mat{D}_E$ defined analogously as $\mat{D}_B$, implying that
\begin{align*}
\lim_{P\rightarrow +\infty} C_s^{\mathrm{OFDM}} 
&= \sum_{m=1}^M \left[\log \frac{|H_B(f_m)|^2}{|H_E(f_m)|^2} \right]^+,\nonumber
\end{align*}
which is equivalent to the secrecy capacity of $M$ parallel independent wiretap channels. For a fixed $L_{CP}$, as $M$ grows large (and thus $N$), the sum converges to an integral and the OFDM secrecy capacity will converge to the generic one.

\section{Secrecy Capacity of FBMC-OQAM}

%

We consider an FBMC-OQAM system with $M$ subcarriers and $N_s$ multicarrier symbols. The real-valued multicarrier symbols, denoted by  $d_{m,l}$ with $m=0,...,M-1$ and $l=0,...,N_s-1$, are modulated using a prototype pulse $g[n]$ of length $M\kappa$, where $\kappa$ is the so-called overlapping factor, \textit{i.e.}, $g[n]=0$ if $n\notin [0,M\kappa-1]$. The FBMC-OQAM modulated signal can be expressed as \cite[Section~2.1]{rottenberg2018fbmc}
\begin{align*}
s[n]=\sum_{m=0}^{M-1}\sum_{l=0}^{N_s-1} d_{m,l} \jmath^{m+l} g[n-lM/2]e^{\jmath \frac{2\pi}{M}m(n-\frac{M\kappa-1}{2})},
\end{align*}
for $n=0,...N-1$ with $N=(N_s+2\kappa-1) M/2$. Note that real-valued multicarrier symbols are spaced only $M/2$ samples apart in time instead of $M+L_{CP}$ for complex symbols in CP-OFDM so that the spectral efficiency is similar for both modulations. To be exact, the CP-OFDM spectral efficiency is penalized by the CP insertion. On the other hand, the FBMC-OQAM spectral efficiency is impacted by tails of length $(2\kappa-1)M/2$ due to the spread of $g[n]$ over multiple multicarrier symbols, which induces an overhead particularly detrimental for small bursts but negligible for long burst (as $N_s$ grows large). Similarly as the capacity analysis in \cite{RezazadehReyhani2017}, we reformulate the transmit signal using a matrix formalism giving
\begin{align*}
	\vect{s}=\mat{A} \tilde{\vect{d}},
\end{align*}
where
\begin{align*}
	\tilde{\vect{d}}= \begin{pmatrix}
	\vect{d}_0\\ \vdots\\ \vect{d}_{N_s-1}
	\end{pmatrix} \in \mathbb{R}^{MN_s\times 1},\ \vect{d}_l= \begin{pmatrix}
	d_{0,l}\\ \vdots\\ d_{M-1,l}
	\end{pmatrix} \in \mathbb{R}^{M\times 1},
\end{align*}
and matrix $\mat{A}\in \mathbb{C}^{N\times MN_s}$ is defined as follows: the element located at the $n$-th row and $m+lM$-th column is given by $\jmath^{m+l} g[n-lM/2]e^{\jmath \frac{2\pi}{M}m(n-\frac{M\kappa-1}{2})}$. At the receiver side, the legitimate receiver, Bob, discards the last $L_B-1$ symbols, using matrix $\mat{S}_B=(\mat{I}_N,\ \mat{0}_{N\times L_B-1})$ and demodulates the signal by applying matrix $\mat{A}^H$. If the filter $g[n]$ has perfect reconstruction properties, this leads to the identity $\Re(\mat{A}^H\mat{A})=\mat{I}_{MN_s}$, so that, under ideal propagation condition and synchronization ($\mat{H}_B=\mat{I}$, $\mat{S}_B=\mat{I}$ and $\vect{w}_B=\vect{0}$), the transmit symbols are recovered after demodulation and real conversion.

In practice however, the multipath channel must be equalized before real conversion to avoid inter-symbol and inter-carrier interference. The obtained samples at Bob and Eve can generally be written as
\begin{align*}
	\vect{z}_B&=\mat{B}_B \mat{H}_B \mat{A} \tilde{\vect{d}} + \mat{B}_B \vect{w}_B,\ \vect{z}_E=\mat{B}_E \mat{H}_E \mat{A} \tilde{\vect{d}} + \mat{B}_E \vect{w}_E.
\end{align*}
The equalizer can be designed in many different ways, see \cite[Section 2.1.5]{rottenberg2018fbmc} for a review. The most conventional way consists in single-tap per-subcarrier equalization as in CP-OFDM. However, as the channel becomes more selective in frequency, the system will be impacted by inter-symbol and inter-carrier interference. Improved equalizer designs rely on a multi-tap structure to estimate the current symbol $d_{m,l}$ based on demodulated symbols at neighboring multicarrier symbols and subcarriers. In the following, we will first derive the FBMC-OQAM secrecy capacity for general matrices $\mat{B}_B$ and $\mat{B}_E$. We will then consider different types of equalizers at Bob and Eve.

Before going further, one should note that vector $\tilde{\vect{d}}$ is real-valued and hence $\mathbb{E}(\tilde{\vect{d}}\tilde{\vect{d}}^T)\neq \mat{0}$ so that vectors $\vect{z}_B$ and $\vect{z}_E$ are improper, \textit{i.e.}, $\mathbb{E}(\vect{z}_B\vect{z}_B^T)\neq \mat{0}$ and $\mathbb{E}(\vect{z}_E\vect{z}_E^T)\neq \mat{0}$. Hence, conventional results for the complex circularly symmetric Gaussian case do not hold. Therefore, we introduce the following real-valued notations for matrices and vectors: for arbitrary vector $\vect{v}$ and matrix $\mat{V}$, vector $\vect{v}_r$ and matrix $\mat{V}_r$ are defined as
\begin{align*}
\vect{v}_{r}=\begin{pmatrix}
\Re(\vect{v})\\ \Im(\vect{v})
\end{pmatrix},\ 	\mat{V}_{r}&=\begin{pmatrix}
\Re(\mat{V}) & -\Im(\mat{V})\\
\Im(\mat{V}) & \Re(\mat{V})
\end{pmatrix}.
\end{align*}
Subscript "$r$" stands for real-valued. We also define $\tilde{\mat{I}}=\begin{pmatrix} \mat{I}_{MN_s}& \mat{0}_{MN_s\times MN_s} \end{pmatrix}^T$. Using these definitions, the perfect reconstruction property becomes $\tilde{\mat{I}}^T \mat{A}_r^T \mat{A}_r \tilde{\mat{I}}=\mat{I}_{MN_s}$ and the power constraint on transmitted symbols $\vect{s}=\mat{A}_r \tilde{\mat{I}}\tilde{\vect{d}}$ directly translates into a constraint on symbols $\tilde{\vect{d}}$
\begin{align*}
\tr\left[\mat{R}_s \right]=\tr\left[\mat{A}_r \tilde{\mat{I}} \mat{R}_{\tilde{d}} \tilde{\mat{I}}^T\mat{A}_r^T\right]=\tr\left[ \mat{R}_{\tilde{d}} \right]\leq P.
\end{align*}
Using this real-valued formalism, the demodulated signal at Bob and Eve can be rewritten as
\begin{align}
\vect{z}_{B,r}&= \mat{B}_{B,r}\mat{H}_{B,r} \mat{A}_r \tilde{\mat{I}} \tilde{\vect{d}} + \mat{B}_{B,r}\vect{w}_{B,r}\nonumber\\
\vect{z}_{E,r}&= \mat{B}_{E,r}\mat{H}_{E,r} \mat{A}_r \tilde{\mat{I}} \tilde{\vect{d}} + \mat{B}_{E,r}\vect{w}_{E,r}.\label{eq:real_MIMO_wiretap_channel}
\end{align}
We are now ready to state our main result.
\begin{theorem} \label{theorem:C_s_multi}
	The FBMC-OQAM secrecy capacity $C_{s}^{\mathrm{FBMC}}$, under a transmit power constraint, can be written as
	\begin{align}
	C_{s}^{\mathrm{FBMC}}&=\max_{\substack{\mat{R}_{\tilde{d}}, \tr\left[\mat{R}_{\tilde{d}}\right]\leq P}}\frac{1}{2} \log \frac{\left| \mat{I} + \mat{T}_{B,r} \mat{R}_{\tilde{d}} \mat{T}_{B,r}^T \right|}{ \left| \mat{I} + \mat{T}_{E,r} \mat{R}_{\tilde{d}} \mat{T}_{E,r}^T \right|},\label{eq:C_s_fbmc}
	\end{align}
	where $\mat{T}_{B,r}=\mat{R}_{w,B}^{-1/2} \mat{B}_{B,r} \mat{H}_{B,r} \mat{A}_r \tilde{\mat{I}}$ and $\mat{T}_{E,r}=\mat{R}_{w,E}^{-1/2}\mat{B}_{E,r}  \mat{H}_{E,r} \mat{A}_r \tilde{\mat{I}}$. Matrices $\mat{R}_{w,B}=1/2 \mat{B}_{B,r}\mat{B}_{B,r}^T$ and $\mat{R}_{w,E}=1/2 \mat{B}_{E,r}\mat{B}_{E,r}^T$ are the noise covariance matrix at Bob and Eve. 
\end{theorem}
\begin{proof}
	See Appendix.
\end{proof}
{\color{black}As in the generic case, this optimization problem is non convex and high dimensional, which makes it hard to solve, even numerically.} The factor $1/2$ comes from the fact that multicarrier symbols composed of real symbols are transmitted instead of complex ones. 
The following result gives a closed-form expression of the secrecy capacity at high SNR. Using again the result of \cite[Th.~2]{Khisti2007} as in the generic and OFDM cases, the FBMC-OQAM secrecy capacity at high SNR is
\begin{align}
	&\lim_{P\rightarrow +\infty} C_{s}^{\mathrm{FBMC}} = \frac{1}{2}\sum_{k=1}^{MN_s}  \left[\log \sigma_k^2\left( \mat{T}_{B,r} \mat{T}_{E,r}^{\dagger} \right) \right]^+.\label{eq:high_SNR_C_s_fbmc}
\end{align}
We now study different equalization structures at Bob and Eve. We distinguish between two extreme cases for Bob receiver structure: 1) multi-tap equalization, 2) single-tap equalization. We consider that Eve also has the choice to use these two receivers plus the generic (ideal) receiver given by $\mat{B}_E=\mat{I}$, which does not discard any received samples and does not apply FBMC-OQAM demodulation.

\subsubsection{Multi-tap Equalization}

In this case, we set $\mat{B}_{B,r}=\mat{A}_r^T \mat{S}_{B,r}$
and we look at the secrecy capacity before real conversion. This can be seen as a multi-tap equalizer making use of the information contained in all of the complex-valued neighboring symbols. Similarly, Eve can also use a multi-tap equalizer $\mat{B}_{E,r}=\mat{A}_r^T \mat{S}_{E,r}$. The FBMC-OQAM secrecy capacity and high SNR secrecy capacity with multi-tap equalization are respectively given by (\ref{eq:C_s_fbmc}) and (\ref{eq:high_SNR_C_s_fbmc}), using the new definition of $\mat{B}_{B,r}$.

\subsubsection{Single-Tap Equalization}

In contrast with the previous section, we now study the performance of the most simple equalization scheme, which consists in i) single-tap equalization by multiplying each subcarrier output by the conjugate of the channel frequency response and ii) real conversion. This gives
\begin{align}
	\mat{B}_{B,r}&=\tilde{\mat{I}}^T \tilde{\mat{D}}_{B,r}^T \mat{A}_r^T \mat{S}_{B,r}, \label{eq:B_single_tap}
\end{align}
where $\tilde{\mat{D}}_{B,r}$ is the real-valued representation of $\tilde{\mat{D}}_{B}=\left(\mat{I}_{N_s} \otimes \mat{D}_{B}\right)$. 
Note that Eve can also use a single-tap equalizer as given in (\ref{eq:B_single_tap}) but matched to her own channel. The FBMC-OQAM secrecy capacity and high SNR secrecy capacity with single-tap equalization are respectively given by (\ref{eq:C_s_fbmc}) and (\ref{eq:high_SNR_C_s_fbmc}), using the new definition of $\mat{B}_{B,r}$ in (\ref{eq:B_single_tap}).

The following theorem shows that, for mildly frequency selective channels, single-tap equalization at Bob incurs no loss as compared to multi-tap equalization and even ideal equalization.
\begin{theorem} \label{theorem:low_frequency_selectivity}
	{\color{black}As $\frac{L_B}{M}\rightarrow 0$ and for well time-frequency localized prototype filters \cite[({As2})]{Rottenberg2017}, FBMC-OQAM demodulation with single-tap and multi-tap equalization achieves the same capacity as ideal equalization. Hence, the secrecy capacity $C_{s}^{\mathrm{FBMC}}$ is identical for single-tap, multi-tap and generic equalization at Bob and Eve, for any $P$. Moreover, the high SNR secrecy capacity then becomes
	\begin{align*}
	&\lim_{P\rightarrow +\infty} C_{s}^{\mathrm{FBMC}} = \frac{N_s}{2} \sum_{m=1}^M \left[\log \frac{|H_B(f_m)|^2}{|H_E(f_m)|^2} \right]^+.
	\end{align*}}
\end{theorem}
\begin{proof}
	See Appendix.
\end{proof}
This result is optimistic in that single-tap equalization has a much lower complexity than multi-tap and generic equalization structures. Moreover, the high SNR secrecy capacity is again equivalent to the secrecy capacity of $M$ parallel independent wiretap channels as in CP-OFDM.

\section{Simulation Results} \label{section:simulation_results}

We now evaluate numerically the high SNR secrecy capacity of the different transceivers	previously studied. Simulations are performed relying on the Matlab-based WaveComBox toolbox \cite{wavecombox}. For a fair comparison, the secrecy capacity is normalized in terms of bits per sample, \textit{i.e.}, dividing obtained expressions of $C_s$ by $N$. The generic secrecy capacity is computed in the large $N$ case. The number of subcarriers and the subcarrier spacing are fixed to $M=64$ and $15$ kHz respectively for both CP-OFDM and FBMC-OQAM modulations. For CP-OFDM, as explained earlier, given that successive multicarrier symbols do not overlap, it is sufficient to consider a single multicarrier symbol, \textit{i.e.}, $N_s=1$. For FBMC-OQAM, the prototype filter is the conventional PHYDYAS pulse \cite{bellanger01} with overlapping factor $\kappa=2$. To take into account the capacity penalty due to tail effects, the number of multicarrier symbols $N_s$ is varied to evaluate differences between short and long burst transmissions.
{\color{black} Bob and Eve channels, $h_B[n]$ and $h_E[n]$, are generated according to either the ITU Veh. A model or the the ITU Veh. B model, \textit{i.e.}, a mildly and a highly frequency selective channel respectively. The averaged power of both channels is normalized to one. }

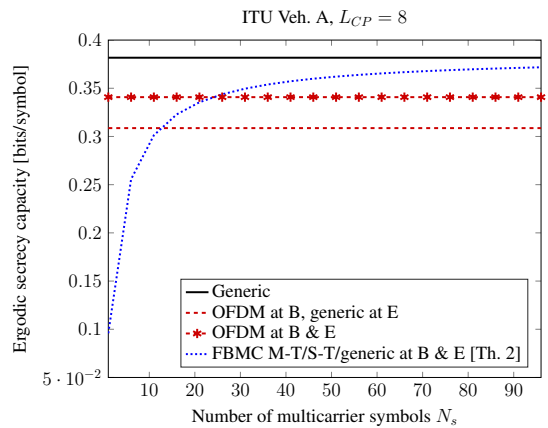
\begin{figure}[!t]
	\centering
	\resizebox{0.4\textwidth}{!}{%
		\Large
%
%
\begin{tikzpicture}

\begin{axis}[%
width=4.520833in,
height=3.527431in,
at={(0.758333in,0.519444in)},
scale only axis,
xmin=1,
xmax=96,
xlabel={Number of multicarrier symbols $N_s$},
ymin=0.05,
ymax=0.4,
ylabel={Ergodic secrecy capacity [bits/symbol]},
title={ITU Veh. A, $L_{CP}=8$},
legend style={at={(0.97,0.03)},anchor=south east,legend cell align=left,align=left,draw=white!15!black}
]
\addplot [color=black,solid,line width=1.5pt]
  table[row sep=crcr]{%
1	0.381694456147186\\
6	0.381694456147186\\
11	0.381694456147186\\
16	0.381694456147186\\
21	0.381694456147186\\
26	0.381694456147186\\
31	0.381694456147186\\
36	0.381694456147186\\
41	0.381694456147186\\
46	0.381694456147186\\
51	0.381694456147186\\
56	0.381694456147186\\
61	0.381694456147186\\
66	0.381694456147186\\
71	0.381694456147186\\
76	0.381694456147186\\
81	0.381694456147186\\
86	0.381694456147186\\
91	0.381694456147186\\
96	0.381694456147186\\
};
\addlegendentry{Generic};

\addplot [color=black!20!red,dashed,line width=1.5pt]
  table[row sep=crcr]{%
1	0.30866984923565\\
6	0.30866984923565\\
11	0.30866984923565\\
16	0.30866984923565\\
21	0.30866984923565\\
26	0.30866984923565\\
31	0.30866984923565\\
36	0.30866984923565\\
41	0.30866984923565\\
46	0.30866984923565\\
51	0.30866984923565\\
56	0.30866984923565\\
61	0.30866984923565\\
66	0.30866984923565\\
71	0.30866984923565\\
76	0.30866984923565\\
81	0.30866984923565\\
86	0.30866984923565\\
91	0.30866984923565\\
96	0.30866984923565\\
};
\addlegendentry{OFDM at B, generic at E};

\addplot [color=black!20!red,dashed,line width=1.5pt,mark size=3.5pt,mark=asterisk,mark options={solid}]
  table[row sep=crcr]{%
1	0.340792340691644\\
6	0.340792340691644\\
11	0.340792340691644\\
16	0.340792340691644\\
21	0.340792340691644\\
26	0.340792340691644\\
31	0.340792340691644\\
36	0.340792340691644\\
41	0.340792340691644\\
46	0.340792340691644\\
51	0.340792340691644\\
56	0.340792340691644\\
61	0.340792340691644\\
66	0.340792340691644\\
71	0.340792340691644\\
76	0.340792340691644\\
81	0.340792340691644\\
86	0.340792340691644\\
91	0.340792340691644\\
96	0.340792340691644\\
};
\addlegendentry{OFDM at B {\&} E};

\addplot [color=blue,dotted,line width=1.5pt]
  table[row sep=crcr]{%
1	0.0958478458195249\\
6	0.255594255518733\\
11	0.301236086861364\\
16	0.322855901707873\\
21	0.335467460368337\\
26	0.343730205697606\\
31	0.349562731812385\\
36	0.353899738410553\\
41	0.357251061690956\\
46	0.359918441444747\\
51	0.362091861984871\\
56	0.363896906162264\\
61	0.365419912186939\\
66	0.366722192700791\\
71	0.36784848936142\\
76	0.368832216824501\\
81	0.36969883387531\\
86	0.370468078223781\\
91	0.371155488067096\\
96	0.371773462572702\\
};
\addlegendentry{FBMC M-T/S-T/generic at B {\&} E [Th. 2]};

\end{axis}
\end{tikzpicture}%
	}
			\vspace{-0.5em}
	\caption{High SNR ergodic secrecy capacity for mildly frequency selective channel.}
	\label{fig:VehA}
	\vspace{-2em}
\end{figure}

{\color{black} Fig.~\ref{fig:VehA} shows the high SNR ergodic secrecy capacity, \textit{i.e.}, the secrecy capacity averaged over channel statistics, of generic, CP-OFDM and FBMC-OQAM modulations as a function of $N_s$, for an ITU Veh. A channel. The gap between the OFDM and the generic secrecy capacity is due to the CP insertion, of length $L_{CP}=M/8=8$ here. Moreover, if Eve is not constrained to use a conventional CP-OFDM receiver, an additional loss is induced. On the FBMC-OQAM side, we see that all transceiver configurations at Bob and Eve, namely, single-tap (S-T), multi-tap (M-T) and generic equalization, reach a similar performance, as foreseen by Th.~\ref{theorem:low_frequency_selectivity} given the mildly frequency selective nature of the ITU Veh. A channel. Moreover, as explained earlier, FBMC-OQAM suffers from a capacity penalty due to tail effects. As $N_s$ grows large, this overhead becomes negligible and it outperforms OFDM.}

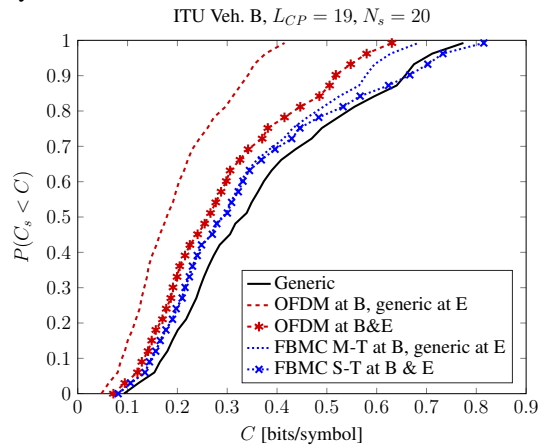
\begin{figure}[!t]
	\centering
	\resizebox{0.4\textwidth}{!}{%
		\Large
%
%
\begin{tikzpicture}

\begin{axis}[%
width=4.520833in,
height=3.565625in,
at={(0.758333in,0.48125in)},
scale only axis,
xmin=0,
xmax=0.9,
xlabel={$C$ [bits/symbol]},
ymin=0,
ymax=1,
ylabel={$P(C_s<C)$},
title={ITU Veh. B, $L_{CP}=19$, $N_s=20$},
legend style={at={(0.97,0.03)},anchor=south east,legend cell align=left,align=left,draw=white!15!black}
]
\addplot [color=black,solid,line width=1.5pt]
  table[row sep=crcr]{%
0.0926236079779542	0\\
0.122002815437361	0.0300751879699248\\
0.153912799944334	0.0601503759398496\\
0.165042423939087	0.0902255639097744\\
0.17990941120598	0.120300751879699\\
0.18977461790846	0.150375939849624\\
0.20089809200941	0.180451127819549\\
0.217495642537782	0.210526315789474\\
0.227691751236787	0.240601503759398\\
0.238046207122874	0.270676691729323\\
0.245313983812754	0.300751879699248\\
0.253394993213892	0.330827067669173\\
0.262604102857177	0.360902255639098\\
0.272913169303503	0.390977443609023\\
0.284817895006517	0.421052631578947\\
0.305655053128822	0.451127819548872\\
0.316263652996557	0.481203007518797\\
0.339064513619241	0.511278195488722\\
0.349050655997946	0.541353383458647\\
0.361790324677377	0.571428571428571\\
0.373746129441794	0.601503759398496\\
0.389036644064258	0.631578947368421\\
0.40842244521199	0.661654135338346\\
0.437130792040296	0.691729323308271\\
0.469402563208006	0.721804511278195\\
0.490120492612495	0.75187969924812\\
0.522704623545615	0.781954887218045\\
0.555489076178998	0.81203007518797\\
0.595925726388642	0.842105263157895\\
0.640136818906466	0.87218045112782\\
0.658231517187397	0.902255639097744\\
0.674917244547057	0.932330827067669\\
0.711543934737734	0.962406015037594\\
0.774162487287076	0.992481203007519\\
};
\addlegendentry{Generic};

\addplot [color=black!20!red,dashed,line width=1.5pt]
  table[row sep=crcr]{%
0.047508184998469	0\\
0.0621963576301266	0.0300751879699248\\
0.0802338675944414	0.0601503759398496\\
0.0859442603515921	0.0902255639097744\\
0.0944581110502675	0.120300751879699\\
0.100311517041481	0.150375939849624\\
0.109646463258868	0.180451127819549\\
0.116725603662251	0.210526315789474\\
0.124189972637221	0.240601503759398\\
0.128156786225192	0.270676691729323\\
0.133598289195531	0.300751879699248\\
0.138982508956785	0.330827067669173\\
0.142433910855186	0.360902255639098\\
0.149164240984227	0.390977443609023\\
0.157952981553704	0.421052631578947\\
0.164936011015008	0.451127819548872\\
0.173716616903323	0.481203007518797\\
0.181412709153406	0.511278195488722\\
0.191048669198252	0.541353383458647\\
0.196953481828579	0.571428571428571\\
0.203622042863265	0.601503759398496\\
0.213451311340151	0.631578947368421\\
0.220777883556336	0.661654135338346\\
0.229876191843925	0.691729323308271\\
0.244258111539002	0.721804511278195\\
0.260473870178942	0.75187969924812\\
0.273427425365825	0.781954887218045\\
0.296599317200814	0.81203007518797\\
0.311374710305463	0.842105263157895\\
0.326769292059771	0.87218045112782\\
0.342143579481469	0.902255639097744\\
0.35425641338119	0.932330827067669\\
0.376821500420928	0.962406015037594\\
0.415106956473779	0.992481203007519\\
};
\addlegendentry{OFDM at B, generic at E};

\addplot [color=black!20!red,dashed,line width=1.5pt,mark size=3.5pt,mark=asterisk,mark options={solid}]
  table[row sep=crcr]{%
0.0715699843603372	0\\
0.0944679503025366	0.0300751879699248\\
0.1198222950354	0.0601503759398496\\
0.128624778434843	0.0902255639097744\\
0.141097060960815	0.120300751879699\\
0.148394634641932	0.150375939849624\\
0.156700897301005	0.180451127819549\\
0.170022930466819	0.210526315789474\\
0.177626912562103	0.240601503759398\\
0.186784907683641	0.270676691729323\\
0.191192350001048	0.300751879699248\\
0.199367372007892	0.330827067669173\\
0.204546011090869	0.360902255639098\\
0.215358765658348	0.390977443609023\\
0.225048802661836	0.421052631578947\\
0.24056914931993	0.451127819548872\\
0.254761920449994	0.481203007518797\\
0.26641525157351	0.511278195488722\\
0.277045534137639	0.541353383458647\\
0.287724590791436	0.571428571428571\\
0.298188642281723	0.601503759398496\\
0.305764174330021	0.631578947368421\\
0.325679549113951	0.661654135338346\\
0.341898287364264	0.691729323308271\\
0.370152222834532	0.721804511278195\\
0.381558168196915	0.75187969924812\\
0.414985340250117	0.781954887218045\\
0.446651186521709	0.81203007518797\\
0.485033883615699	0.842105263157895\\
0.505422863052472	0.87218045112782\\
0.51809874210679	0.902255639097744\\
0.547912508694315	0.932330827067669\\
0.579886995074109	0.962406015037594\\
0.630445860533763	0.992481203007519\\
};
\addlegendentry{OFDM at B{\&}E};

\addplot [color=blue,dotted,line width=1.5pt]
  table[row sep=crcr]{%
0.0810604960134092	0\\
0.106717590581859	0.0300751879699248\\
0.134882188768874	0.0601503759398496\\
0.144371643251624	0.0902255639097744\\
0.157449193009238	0.120300751879699\\
0.165808330528255	0.150375939849624\\
0.17589279391278	0.180451127819549\\
0.190463141421747	0.210526315789474\\
0.19909664094668	0.240601503759398\\
0.208913875269181	0.270676691729323\\
0.214892080850967	0.300751879699248\\
0.222377580860586	0.330827067669173\\
0.230347157918296	0.360902255639098\\
0.239454589055202	0.390977443609023\\
0.249953452173931	0.421052631578947\\
0.268735755535717	0.451127819548872\\
0.277290878430035	0.481203007518797\\
0.297984221132795	0.511278195488722\\
0.306828844914096	0.541353383458647\\
0.318021849707632	0.571428571428571\\
0.32751658149917	0.601503759398496\\
0.341232449050251	0.631578947368421\\
0.359614122460087	0.661654135338346\\
0.385085613347325	0.691729323308271\\
0.413250017351687	0.721804511278195\\
0.432000377994452	0.75187969924812\\
0.460977167425045	0.781954887218045\\
0.493150416368925	0.81203007518797\\
0.525381240036369	0.842105263157895\\
0.564736222499331	0.87218045112782\\
0.579105302762134	0.902255639097744\\
0.595769312161689	0.932330827067669\\
0.628124467565078	0.962406015037594\\
0.683771531205785	0.992481203007519\\
};
\addlegendentry{FBMC M-T at B, generic at E};

\addplot [color=blue,dotted,line width=1.5pt,mark size=3.5pt,mark=x,mark options={solid}]
  table[row sep=crcr]{%
0.0809423068107854	0\\
0.105979183190861	0.0300751879699248\\
0.134857837805419	0.0601503759398496\\
0.143904626960734	0.0902255639097744\\
0.156503669798653	0.120300751879699\\
0.165554131087904	0.150375939849624\\
0.176974277004985	0.180451127819549\\
0.190199331723399	0.210526315789474\\
0.197604897204977	0.240601503759398\\
0.20962983238401	0.270676691729323\\
0.215360898912944	0.300751879699248\\
0.224098718848987	0.330827067669173\\
0.230273233256345	0.360902255639098\\
0.2398376139903	0.390977443609023\\
0.248375402731804	0.421052631578947\\
0.270152787281118	0.451127819548872\\
0.279168651121874	0.481203007518797\\
0.300283701836699	0.511278195488722\\
0.30880010709173	0.541353383458647\\
0.322210526582796	0.571428571428571\\
0.33170930073443	0.601503759398496\\
0.34472441572403	0.631578947368421\\
0.368515224638411	0.661654135338346\\
0.396009654072104	0.691729323308271\\
0.429512761981277	0.721804511278195\\
0.445944320163693	0.75187969924812\\
0.483758325285571	0.781954887218045\\
0.53263984269862	0.81203007518797\\
0.566889963013284	0.842105263157895\\
0.623939930213425	0.87218045112782\\
0.666548899396531	0.902255639097744\\
0.70261856238858	0.932330827067669\\
0.732912318222255	0.962406015037594\\
0.815072837200117	0.992481203007519\\
};
\addlegendentry{FBMC S-T at B {\&} E};

\end{axis}
\end{tikzpicture}%
	}
			\vspace{-0.5em}
	\caption{Cumulative density function of the high SNR secrecy capacity for highly frequency selective channel.}
	\label{fig:VehB}
	\vspace{-2em}
\end{figure}

{\color{black}Fig.~\ref{fig:VehB} plots the cumulative density function of the high SNR secrecy capacity for the highly frequency selective ITU Veh. B channel. The OFDM gap from the generic secrecy capacity is increased because the CP length has to be increased to $L_{CP}=19$ to compensate for the longer channel impulse response. For a fixed number of multicarrier symbols $N_s=20$, the FBMC curves outperform OFDM ones. We also see that the secrecy capacity is improved if Eve uses FBMC-OQAM demodulation and single-tap equalization. In some cases, the secrecy capacity becomes even higher than the generic secrecy capacity. Note that this is only possible because Eve uses a suboptimal receiver.}

\section{Conclusion} \label{section:conclusion}
In this paper, we have characterized the FBMC-OQAM secrecy capacity over a frequency selective channel. The secrecy capacity is formulated as an optimization problem that has a closed-form in the high SNR regime. Single-tap and multi-tap equalizers were compared and were shown to be equivalent for mildly frequency selective channels. We have also shown that FBMC-OQAM is particularly competitive for medium-to-long burst transmission as compared to the OFDM and generic secrecy capacity. {\color{black} A promising research direction includes the extension of this study to multiple-antenna systems taking different FBMC-OQAM beamforming and equalization techniques into account.}

\section{Appendix} \label{section:appendix}


\textbf{Proof of Theorem~\ref{theorem:C_s_multi}}:
We need to derive the secrecy capacity of the real MIMO wiretap channel given in (\ref{eq:real_MIMO_wiretap_channel}),
under the transmit power constraint $\tr\left[\mat{R}_s \right]=\tr\left[ \mat{R}_{\tilde{d}} \right]\leq P$. The noise samples can be colored after demodulation and equalization, depending on the structure of $\mat{B}_{B,r}$ and $\mat{B}_{E,r}$. To whiten the noise samples, one can multiply $\vect{z}_{B,r}$ and $\vect{z}_{E,r}$ by $\mat{R}_{w,B}^{-1/2}$ and $\mat{R}_{w,E}^{-1/2}$ respectively, where $\mat{R}_{w,B}$ and $\mat{R}_{w,E}$ are defined in Th.~\ref{theorem:C_s_multi}. Note that this operation is invertible and does not affect the information contained at Bob and Eve. Using the definitions introduced in Th.~\ref{theorem:C_s_multi}, we obtain
\begin{align*}
\tilde{\vect{z}}_{B,r}&= \mat{T}_{B,r} \tilde{\vect{d}} + \tilde{\vect{w}}_{B,r},\ \tilde{\vect{z}}_{E,r}= \mat{T}_{E,r} \tilde{\vect{d}} + \tilde{\vect{w}}_{E,r}.
\end{align*}
We can then apply the result of \cite[Th.~3]{Liu2009} and we find the result of Th.~\ref{theorem:C_s_multi}. Secrecy capacity is achieved without channel prefixing and by choosing $\tilde{\vect{d}}$ as a zero mean real Gaussian vector with covariance $\mat{R}_{\tilde{d}}$. 


\textbf{Proof of Theorem~\ref{theorem:low_frequency_selectivity}}:
We need to show that the capacity with single-tap, multi-tap and ideal equalization is equivalent for mildly frequency selective channels. For clarity we omit subscripts "$B$" and "$E$" in this section as the result needs to be proven at Bob and Eve and the proof is completely symmetrical. Given the identity $|\mat{I}+\mat{A}\mat{B}|=|\mat{I}+\mat{B}\mat{A}|$, the FBMC-OQAM capacity with single-tap, multi-tap and ideal equalization become equivalent if the product $(\mat{T}_{r})^T\mat{T}_{r}=\mat{K}$ for some fixed $\mat{K}$ and for $\mat{T}_{r}\in \{\mat{T}_{r}^{\mathrm{Single}},\mat{T}_{r}^{\mathrm{Multi}},\mat{T}_{r}^{\mathrm{Gen}}\}$. To show this, we will use the three following results. Under general assumptions on the prototype filter $g[n]$, the methodology used in \cite{Rottenberg2017}, relying on a Taylor approximation of the channel variations in frequency, can be used to show that 
\begin{align}
\mat{A}_r^T \mat{S}_{r} \mat{H}_{r} \mat{A}_r&= \mat{A}_r^T  \mat{A}_r \tilde{\mat{D}}_{r}+\vect{\epsilon}_1
\label{eq:first_identity_channel_selectivity}\\
\tilde{\mat{I}}^T \tilde{\mat{D}}_{r}^T   \mat{A}_r^T \mat{A}_r \tilde{\mat{D}}_{r} \tilde{\mat{I}}
&= \left(\mat{I}_{N_s} \otimes \mat{D}^H\mat{D}\right)+\vect{\epsilon}_2 \label{eq:second_identity_channel_selectivity}\\
\tilde{\mat{I}}^T\mat{A}_r^T \mat{H}_{r}^T \mat{H}_{r} \mat{A}_r\tilde{\mat{I}}&= \left(\mat{I}_{N_s} \otimes \mat{D}^H\mat{D}\right)+\vect{\epsilon}_3, \label{eq:third_identity_channel_selectivity}
\end{align}
and the approximation errors $\|\vect{\epsilon}_1\|$, $\|\vect{\epsilon}_2\|$ and $\|\vect{\epsilon}_3\|$ go to zero at rate $\frac{L}{M}$ as $\frac{L}{M}\rightarrow 0$. For the multi-tap case, we have $\mat{B}_{r}=\mat{A}_r^T \mat{S}_{r}$ and 
\begin{align*}
\lim_{{L}/{M}\rightarrow 0} \mat{T}_{r}^{\mathrm{Multi}}
&\stackrel{(\ref{eq:first_identity_channel_selectivity})}{=} \mat{R}_{w,B}^{-1/2} \mat{A}_r^T \mat{A}_r \mat{D}_{r} \tilde{\mat{I}}
\end{align*}
\begin{align*}
\lim_{{L}/{M}\rightarrow 0} \left(\mat{T}_{r}^{\mathrm{Multi}}\right)^T  \mat{T}_{r}^{\mathrm{Multi}} &\stackrel{(\ref{eq:second_identity_channel_selectivity})}{=} 2 \left(\mat{I}_{N_s} \otimes \mat{D}^H\mat{D}\right).
\end{align*}
For the single-tap case, we have $\mat{B}_{r}=\tilde{\mat{I}}^T \tilde{\mat{D}}_{r}^T \mat{A}_r^T \mat{S}_{r}$ and
\begin{align*}
\lim_{{L}/{M}\rightarrow 0}  \mat{T}_{r}^{\mathrm{Single}}
&\stackrel{(\ref{eq:first_identity_channel_selectivity},\ref{eq:second_identity_channel_selectivity})}{=} \mat{R}_{w,B}^{-1/2}\left(\mat{I}_{N_s} \otimes \mat{D}^H\mat{D}\right)\\
\lim_{{L}/{M}\rightarrow 0} (\mat{T}_{r}^{\mathrm{Single}})^T\mat{T}_{r}^{\mathrm{Single}}
&\stackrel{(\ref{eq:second_identity_channel_selectivity})}{=} 2 \left(\mat{I}_{N_s} \otimes \mat{D}^H\mat{D}\right).
\end{align*}
For the generic (ideal) case, we have $\mat{B}_{r}=\mat{I}$ and
\begin{align*}
\mat{T}_{r}^{\mathrm{Gen}}&=\sqrt{2}\mat{H}_{r} \mat{A}_r\tilde{\mat{I}}\\
\lim_{{L}/{M}\rightarrow 0}  (\mat{T}_{r}^{\mathrm{Gen}})^T\mat{T}_{r}^{\mathrm{Gen}}&\stackrel{(\ref{eq:third_identity_channel_selectivity})}{=} 2 \left(\mat{I}_{N_s} \otimes \mat{D}^H\mat{D}\right),
\end{align*}
which shows that $(\mat{T}_{r})^T\mat{T}_{r}$ is well identical for each type of equalization structure. Inserting these last results in (\ref{eq:high_SNR_C_s_fbmc}) 
gives the final result of Th.~\ref{theorem:low_frequency_selectivity}.

%

\footnotesize
\bibliographystyle{IEEEtran}
\bibliography{IEEEabrv,IEEEreferences}

\end{document}